%% file: ecir_preprint.tex
\documentclass[runningheads]{llncs}
\usepackage[T1]{fontenc}
\usepackage{amsmath}
\usepackage{amsfonts}
\usepackage{amssymb}
\usepackage{bm}
\usepackage{multirow}
\usepackage{multicol}
\usepackage{booktabs}
\usepackage{graphicx}
\usepackage{float}
\usepackage{hyperref}
\usepackage{xcolor}
\usepackage{algorithm}
\usepackage{algorithmicx}
\usepackage{algpseudocode}

\usepackage[most,skins,theorems]{tcolorbox}
\tcbuselibrary{breakable}

\tcbset{
  aibox/.style={
    width=\linewidth,
    top=8pt,
    bottom=4pt,
    colback=blue!6!white,
    colframe=black,
    colbacktitle=black,
    enhanced,
    center,
    attach boxed title to top left={yshift=-0.1in,xshift=0.15in},
    boxed title style={boxrule=0pt,colframe=white,},
  }
}

\usepackage{pifont}

\begin{document}
\title{Adapting Generative Recommenders for Multi-Turn Interaction}
%
%
\author{Yu-Chen Den\inst{1} \and Zhi Rui Tam\inst{1} \and Yung-Yu Shih\inst{1} \and Shih-Hsin Wang\inst{1} \and Yun-Nung Chen\inst{1} \and Pu-Jen Cheng\inst{1} \and Eugene Yang\inst{2}}

\authorrunning{Den et al.}
\institute{National Taiwan University \email{\texttt{\{d13944011,d14922019,f12944007,shwang\}}@csie.ntu.edu.tw\\y.v.chen@ieee.org, pjcheng@csie.ntu.edu.tw} \and Johns Hopkins University \\ \email{eugene.yang@jhu.edu}}

%
%
\maketitle              

\begin{abstract}
Generative recommenders decode items from a user's interaction history, but offer no way for users to correct a recommendation that misses their current intent. Adding conversation is natural since items and words share same output space, yet training the model to converse may overwrite the history-to-item mapping it relies on. We introduce INTEGER (\textbf{INTE}ractive \textbf{GE}nerative \textbf{R}ecommendation), which extends generative recommendation to multi-turn interaction with a learned routing token that lets the model decide when to recommend, history re-anchoring that conditions each item on both past behavior and the dialogue, and behavioral replay with instruction-data rehearsal that prevents forgetting during adaptation. Users can thus give feedback on recommendations within the dialogue, while recommendations stay grounded in behavioral history and accuracy is not traded for fluency. On Amazon Beauty and Toys, INTEGER matches or exceeds the strongest baselines in accuracy with competitive conversation quality, improving Hit@10 by 13.3\% on Amazon Beauty, and significantly outperforms the generative recommender it starts from. Our analyses show that INTEGER learns behaviors that naive adaptation fails to acquire, recommending once the user's intent is clear and staying attentive to behavioral history at the moment of recommendation. INTEGER also learns an intent-agnostic replacement over the item space, which suppresses rejected items but points to attribute-aware feedback as the next step.

\end{abstract}

\input{acl_paper/intro}
\input{acl_paper/background}
\input{acl_paper/method}
\input{acl_paper/experiment}
\input{acl_paper/discussions}
\input{acl_paper/conclusions_and_future_works}

\bibliographystyle{splncs04}
\bibliography{ecir_preprint}
\end{document}

%% file: acl_paper/intro.tex
\section{Introduction}
Generative recommendation has recently emerged as a promising paradigm for sequential recommendation~\cite{deng2025onerec,han2025mtgr,he2026plum,rajput2023recommender,zhai2024actions}. Rather than retrieving items from an external index built over learned user and item embeddings~\cite{hidasi2015session,kang2018self,sun2019bert4rec}, generative recommendation decodes the identifier of the target item directly, representing each item as a sequence of discrete codes derived from its content embedding~\cite{jegou2010product,lee2022autoregressive}. This places items and words in the same output space and has proven competitive with other strong sequential recommenders~\cite{kang2018self,kim2025lost,sun2019bert4rec}.

However, these models rely on a fixed interaction history to infer user preferences and generate recommended items, giving users no direct way to correct the system when its recommendation misses their current intent~\cite{rajput2023recommender,wang2025cesrec,zhai2024actions}. When the user's preference changes or the recommendation fails to capture what they currently want, the model cannot receive that correction explicitly. Instead, the updated preference must first be reflected in subsequent item interactions before it becomes observable to the model, making preference correction indirect and potentially slow.

Conversational recommendation systems (CRS) let users express preferences through dialogue~\cite{ravaut2024parameter,wang2022towards}, but they typically learn to recommend from the conversational context itself. Generative recommenders, in contrast, learn a direct mapping from interaction history to item identifiers, capturing preference signals accumulated from users' past actions. When item identifiers are tokens in a language model's vocabulary, recommending an item and generating a response become the same operation, so conversational ability can be added without architectural change. However, this shared output space also means that both behaviors are carried by the same parameters: training the model to converse may overwrite the history-to-item mapping it has already acquired~\cite{verma2026orbit}. We therefore ask whether \emph{\textbf{a generative recommender can learn to converse without losing the recommendation behavior it was originally trained on.}}

We solve this problem with \emph{\textbf{INTE}ractive \textbf{GE}nerative \textbf{R}ecommendation (INTEGER)}, which empowers generative recommendation to interact with users through multi-turn conversation. Its design is grounded in a single principle: conversation acts as a control layer over a generative recommender instead of recommendation signals. Recommendation therefore remains in the history-to-item mapping acquired from interaction sequences, while dialogue governs \emph{when} that mapping is invoked and \emph{how} its output is presented and revised.

Our contributions are threefold. First, we expose recommendation as a learned decoding action, where a routing token decides when to recommend and \emph{history re-anchoring} re-emits the user history right before item generation. Second, we combine behavioral replay with instruction-data rehearsal to preserve recommendation accuracy during conversational adaptation. On Amazon Beauty and Toys~\cite{hou2026bridging}, INTEGER is competitive with both end-to-end CRS and modular pipelines and significantly outperforms the generative recommender it is adapted from. Third, our analyses show how conversational and behavioral training erode each other and how re-anchoring and rehearsal counteract this, and that refinement learns a single intent-agnostic replacement. Dataset and code will be released after the paper is accepted.

%% file: acl_paper/background.tex
\section{Background}
\subsection{Multi-Turn Interaction}
\label{prelim:conv}
In multi-turn conversation, a model converses with a user over multiple turns. We write the dialogue up to the \(t\)-th turn as \(\mathcal{D}_t = (u_1, a_1, \dots, u_{t-1}, a_{t-1}, u_t)\), where \(u_i\) and \(a_i\) denote the \(i\)-th user message and assistant response, and \(\mathcal{D}_t\) ends at the pending user message \(u_t\). The response \(a_t\) is generated autoregressively conditioned on \(\mathcal{D}_t\). When the interaction serves recommendation, the model needs to identify the turns at which a recommendation is warranted and surface an item consistent with the intent expressed so far in $\mathcal{D}_t$, while still generating fluent responses.

Existing work approaches this task in three main ways. Conversational recommender systems (CRS) learn recommendation directly from multi-turn dialogue, jointly modeling preference elicitation, item recommendation, and response generation~\cite{ravaut2024parameter,wang2022towards,zhang2026generative}. Other work treats conversation as an additional feedback channel for an existing recommender, converting it into pseudo-interactions for sequential recommendation~\cite{wang2025cesrec} or into preference signals~\cite{tajiri2025refining}. A third line adopts a modular design, in which an LLM handles the interaction and delegates recommendation to an external model or tool~\cite{gao2023chatrec,huang2025interecagent,li2025chatcrs,zhao2024toolrec}.

\subsection{Generative Recommendation}
\label{prelim:genrec}

Conventional sequential recommenders~\cite{hidasi2015session,kang2018self,ma2019hierarchical,sun2019bert4rec} assign each item a unique ID and score items through a large embedding table. Generative recommendation instead represents each item as a sequence of discrete tokens and generates the next item by autoregressive decoding~\cite{deng2025onerec,he2026plum,rajput2023recommender,zhai2024actions}.
Existing work has focused on semantic ID design~\cite{ju2025generative,liu2025generative,zhu2024cost}, training and decoding algorithms~\cite{han2025mtgr,hou2025generating,lin2026bringing,wang2026nezha,zheng2025pre}, and multimodal item representations~\cite{liu2024mmgrec,vandenhirtz2026multimodal,xu2026mmq,zhu2025beyond}.
In all of these settings, recommendation is one-shot: the model produces an item from a fixed interaction history, and the user cannot revise its input except through further interactions.

Formally, a generative recommender maps a user's interaction history \(H = (h_1, \dots, h_T)\) to a ranked list of item identifiers (ItemIDs). Each item is represented by an ItemID \(\bm{z} = (c_1, \dots, c_L)\), a sequence of \(L\) discrete codewords obtained by RQ-KMeans~\cite{ju2025generative} from the item's content embedding, so that semantically similar items share longer codeword prefixes. The model \(p_{\theta}\) assigns each ItemID a likelihood autoregressively:
\begin{equation}
    p_{\theta}\left(\bm{z} \mid H\right)
    = \prod_{\ell=1}^{L} p_{\theta}\!\left(c_{\ell} \mid \bm{c}_{<\ell}, H\right),
    \label{eq:retrieval_gr}
\end{equation}
\noindent where \(\bm{c}_{<\ell} = (c_1, \dots, c_{\ell-1})\) denotes the prefix preceding position \(\ell\). At inference time, the top-\(N\) list \(\mathcal{Z} = \left(\bm{z}^{(1)}, \dots, \bm{z}^{(N)}\right)\) is obtained by prefix-trie-constrained beam search over this factorization, with \(p_{\theta}(\bm{z}^{(1)} \mid H) \geq \cdots \geq p_{\theta}(\bm{z}^{(N)} \mid H)\).
The trie constraint restricts decoding to prefixes of valid ItemIDs, ensuring every generated sequence corresponds to an item in the corpus. 

INTEGER starts from such a model and adapts it to the conversational task in Section~\ref{prelim:conv}. As in end-to-end CRS, a single model both converses and recommends; unlike CRS, its recommendation ability comes from the history-to-item mapping learned on interaction sequences, with conversation acting as a control layer over that mapping. Unlike modular pipelines, it requires no separate recommender.

%% file: acl_paper/method.tex
\section{INTEGER: Interactive Generative Recommendation}
\label{sec:method}

\begin{figure*}[!t]
    \centering
    \includegraphics[width=\linewidth]{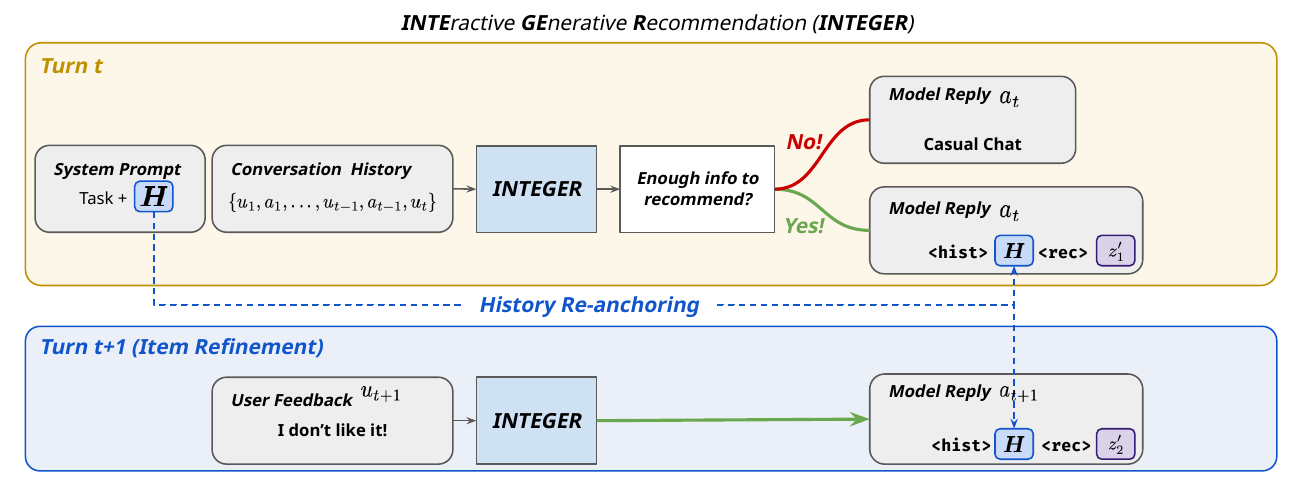}
    \caption{Overview of INTEGER. Starting from a generative recommender trained on interaction sequences, INTEGER uses a learned routing token to choose between chatting and recommending. When recommending, it places the user's behavioral history immediately before ItemID generation. Subsequent user feedback can trigger a revised recommendation.}
    \label{fig:model_struct}
\end{figure*}

We start with a generative recommender~\cite{rajput2023recommender} trained to predict an ItemID from a user's interaction history \(H\) (Section~\ref{prelim:genrec}). To apply this model to conversation, we place \(H\) in the system prompt at the beginning of the dialogue. At each turn \(t\), the model receives this prompt and the dialogue context \(\mathcal{D}_t\), then decides whether to continue chatting or recommend an item. INTEGER makes recommendation a decoding action and places the interaction history next to the generated ItemID when that action is taken. Figure~\ref{fig:model_struct} illustrates the inference process.

\subsection{Recommendation as a Decoding Action}
\label{sec:decoding-action}

We introduce a learned routing token \texttt{<hist>} into the model's vocabulary. At the start of an assistant turn, emitting \texttt{<hist>} initiates a recommendation; otherwise, the model produces a conversational response. The decision to recommend is therefore made by the same model that generates the response, without an external routing classifier~\cite{chen2025llm,pattnayak2025hybrid}.

Internally, a recommendation turn follows the format \(\texttt{<hist>}\; H \;\texttt{<rec>}\; \bm{z}^{\prime}\), where \(\bm{z}^{\prime}\) is the recommended item's ItemID. After the model emits \texttt{<hist>}, the inference procedure inserts \(H\) and \texttt{<rec>} into the decoding context. The model then generates \(\bm{z}^{\prime}\) under a prefix-trie constraint, which restricts decoding to valid ItemIDs. Once the ItemID is complete, it resumes unconstrained decoding to produce the accompanying response.

We call the insertion of \(H\) at inference time \emph{history re-anchoring}. The copy of \(H\) in the system prompt remains available while the model interprets the dialogue and decides whether to recommend. Re-anchoring places the same history immediately before ItemID generation, recovering the local history-to-item format used to train the original generative recommender. Item generation can thus draw on the behavioral history together with preferences expressed in the conversation. Algorithm~\ref{alg:decode} specifies the decoding procedure.

\begin{algorithm}[t]
\caption{Routing-aware constrained decoding}
\label{alg:decode}
\begin{algorithmic}[1]
\Require System prompt \(M\); dialogue context \(\mathcal{D}_t\); model \(p_\theta\); catalog trie \(\mathcal{T}\) over ItemID codewords; structural-token set \(\mathcal{K}\)
\State \(\bm{p} \gets \log p_\theta(\cdot \mid M,\mathcal{D}_t)\) \Comment{next-token log-probability vector}
\State \(\bm{p}[\mathcal{K}\setminus\{\texttt{<hist>}\}] \gets -\infty\)
\State \(v^* \gets \arg\max_v \bm{p}[v]\)
\If{\(v^*=\texttt{<hist>}\)} \Comment{recommend}
    \State Insert \(H\;\texttt{<rec>}\) after \texttt{<hist>}
    \State Decode \(\bm{z}^{\prime}\) under trie \(\mathcal{T}\)
    \State Resume unconstrained response decoding until \texttt{<|im\_end|>}
\Else \Comment{chat}
    \State Append \(v^*\) and continue unconstrained decoding until \texttt{<|im\_end|>}
\EndIf
\end{algorithmic}
\end{algorithm}

\subsection{Conversational Adaptation with Behavioral Replay}
\label{sec:recipe}

We adapt the generative recommender with a mixture of three training sources: multi-turn recommendation dialogues, sequential recommendation examples, and general instruction-following examples. Each source teaches a different part of the desired behavior.

The multi-turn dialogues teach the model when to emit \texttt{<hist>}, how to respond conversationally, and how to produce recommendations in the format shown in Section~\ref{sec:decoding-action}. Training only on these dialogues can weaken the history-to-item mapping acquired during generative recommendation training~\cite{verma2026orbit}. We therefore reuse examples from the original sequential recommendation training set during adaptation. We call this \emph{behavioral replay}: the model continues to train on history-to-ItemID prediction while learning to converse. Behavioral replay acts on the model parameters during training, whereas history re-anchoring places \(H\) in the decoding context when a recommendation is generated.

We also include general instruction-following examples~\cite{ding2023enhancing} to strengthen response generation alongside the recommendation-focused training data. Together, the three sources train the model to manage a dialogue while retaining the behavioral mapping used for item generation. We train on their mixture with the standard next-token prediction loss:
\begin{equation}
    \mathcal{L}_{\mathrm{SFT}}(\theta)
    =
    -\mathbb{E}_{(x,y)\sim\mathcal{D}_{\mathrm{mix}}}
    \sum_{i=1}^{|y|}
    \log p_{\theta}(y_i\mid x,y_{<i}),
    \label{eq:conv_loss}
\end{equation}
where \(\mathcal{D}_{\mathrm{mix}}\) combines the three sources, \(x\) is the context provided to the model, and \(y\) is the target sequence on which the loss is computed. In recommendation turns, the re-anchored history \(H\) and \texttt{<rec>} are excluded from the loss, we supervise only \texttt{<hist>}, the ItemID \(\bm{z}^{\prime}\), and the subsequent response.

\subsection{Item Refinement}
\label{sec:refinement}

After receiving a recommendation, a user may reject it or ask for another item similar to one they accepted. INTEGER handles these follow-up requests within the same dialogue: the previous recommendation and the user's feedback become part of \(\mathcal{D}_{t+1}\), which let the routing and ItemID decoding procedure can produce a revised recommendation.

We train this behavior using preference pairs constructed for two feedback types. For \emph{rejection}, \(y^-\) is the model's erroneous top-1 recommendation and \(y^+\) is the ground-truth item. For \emph{variety}, \(y^-\) is the currently recommended item and \(y^+\) is its nearest neighbor, determined by cosine similarity between item content embeddings. Both pair types share the same structure: the preferred continuation replaces the current item rather than repeating it. We further apply Direct Preference Optimization (DPO)~\cite{rafailov2023direct}:
\begin{equation}
    \mathcal{L}_{\mathrm{DPO}}(\theta)
    =
    -\log \sigma\!\left[
        \beta\left(
            \log\frac{\pi_\theta(y^+\mid x)}
                     {\pi_{\mathrm{ref}}(y^+\mid x)}
            -
            \log\frac{\pi_\theta(y^-\mid x)}
                     {\pi_{\mathrm{ref}}(y^-\mid x)}
        \right)
    \right],
    \label{eq:dpo}
\end{equation}
where \(x\) contains the history prompt and the dialogue of the user's feedback, \(\pi_{\mathrm{ref}}\) is the model before DPO, and \(\beta\) controls the strength of the preference update.

%% file: acl_paper/experiment.tex
\section{Experiments}

\subsection{Experimental Settings}

\paragraph{\textbf{Datasets.}} We evaluate INTEGER on two public benchmarks derived from the Amazon Product Reviews dataset~\cite{hou2026bridging}, adopting the ``Beauty'' and ``Toys and Games'' categories following common practice in sequential recommendation~\cite{rajput2023recommender,yang2026sparse}. Since no large-scale dataset pairs behavioral histories with multi-turn dialogue, we synthesize conversations for both categories using Tencent-Hy3-preview~\cite{tencent2025hy3preview}. The statistics are summarized in Table~\ref{tab:data}.

\input{table/data_stats}

\paragraph{\textbf{Baselines.}} We compare against three groups of baselines. The first consists of sequential recommenders, SASRec~\cite{kang2018self} and TIGER~\cite{rajput2023recommender}, which provide recommendation-only reference points; TIGER is built on Qwen3-1.7B-Base~\cite{yang2025qwen3} and is the model INTEGER starts from. The second consists of end-to-end conversational recommenders, GCRS~\cite{zhang2026generative} and PECRS~\cite{ravaut2024parameter}. The third consists of modular pipelines that pair the same TIGER recommender with a Qwen3-1.7B generator: Modular (Classic) always retrieves and verbalizes a candidate list, while Modular (Tool-call) lets the generator decide when to invoke the recommender, isolating the benefit of learning the recommendation action rather than delegating it.

\paragraph{\textbf{Evaluation Metrics and Statistical Tests.}} We report Hits@10 and NDCG@10 for recommendation accuracy. For conversational systems, both are computed at the ground-truth recommendation turn of each test dialogue; if a system that decides when to recommend (INTEGER or Modular (Tool-call)) does not recommend at this turn, the turn is counted as a miss, so routing errors are reflected in retrieval performance. For conversation quality, we avoid \(n\)-gram overlap metrics such as BLEU~\cite{papineni2002bleu} and ROUGE~\cite{lin2004rouge}, which correlate poorly with human judgment. Instead, we adopt an LLM-as-judge protocol in which DeepSeek-V4-Flash~\cite{xu2026deepseek} rates each response from 1 to 5 on naturalness, relevance, helpfulness, and coherence, and we report the mean as Conv.~Q~\cite{doh2026llm,zheng2023judging}.

We use paired \(t\)-tests against a reference baseline. For accuracy, the reference is TIGER, from which INTEGER is adapted, paired at the user level. For Conv.~Q, the reference is Modular (Tool-call), the strongest conversational baseline on Beauty and the only other one that decides when to recommend, paired at the turn level.

\paragraph{\textbf{Implementation Details.}} Table~\ref{tab:hparams} lists the hyperparameters for training. We use Axolotl\footnote{\url{https://axolotl.ai/}} for finetuning models in all stages. For semantic ID construction, items are embedded with \texttt{sentence-t5-base}~\cite{ni2022sentence} and residual-quantized into three levels of \(256\) codes, with the fourth level for disambiguity.
 
\begin{table}[t]
\centering
\caption{Training hyperparameters. Effective batch size is
micro-batch $\times$ gradient accumulation. Stage-1 denotes training generative recommendation and Stage-2 denotes our conversational tuning.}
\label{tab:hparams}
\setlength{\tabcolsep}{5pt}
\begin{tabular}{lccc}
\toprule
 & \textbf{Stage-1} & \textbf{Stage-2} & \textbf{DPO} \\
\midrule
Initialized from & Qwen3-1.7B-Base~\cite{yang2025qwen3} & Stage-1 ckpt & Stage-2 ckpt \\
Learning rate & $5\times10^{-4}$ & $1\times10^{-4}$ & $5\times10^{-6}$ \\
Effective batch & $8\times32=256$ & $8\times16=128$ & $2\times8=16$ \\
Budget & $10$ epochs & $5$ epochs & $1$ epoch \\
Weight decay & $10^{-6}$ & $10^{-6}$ & $0$ \\
Warmup ratio & 0.1 & 0.1 & 0.1 \\
RL \(\beta\) & -- & -- & 0.1 \\
\bottomrule
\end{tabular}
\end{table}

\subsection{Main Results}
\input{table/main}

Table~\ref{tab:main_result} summarizes results on both categories, where the baselines fail in two opposite ways. GCRS attains near-zero accuracy (Hits@10 of 0.004 and 0.002) despite competitive conversation quality, while PECRS occupies the opposite, reaching the strongest baseline NDCG@10 on Toys, but at the lowest conversation quality of all systems (3.92 and 4.05). The modular pipelines sit between these extremes, recovering fluency (4.40--4.58) at moderate accuracy.

On Beauty, INTEGER obtains the highest score on all three reported measures: its Hits@10 increases from the best baseline value of 0.045 to 0.051, and its NDCG@10 from 0.025 to 0.027, and conversation quality from 4.58 to 4.77, the last significantly above Modular (Tool-call) (paired \(t\)-test, \(p<0.001\)). On Toys, INTEGER ties PECRS for the best Hits@10 (0.057) with 0.20 points higher in conversation quality, while PECRS retains a higher NDCG@10 (0.032 vs.\ 0.028) and the modular pipelines receive higher conversation quality. On both datasets, INTEGER significantly outperforms TIGER, the generative recommender it is adapted from, in Hits@10 and NDCG@10 (paired \(t\)-test, \(p<0.001\)). Conversational adaptation therefore does not erode the history-to-item mapping; combined with the preferences expressed in the dialogue, it raises accuracy above what the interaction history alone supports.

\subsection{Ablation Studies}
\input{table/ablation}

Table~\ref{tab:ablation} confirms that each component contributes to at least recommendation performance or conversation quality and that the two behavioral safeguards carry most of the weight. Training on dialogues alone collapses on both axes (Hits@10=$0.029$, Conv. Q=$2.39$). Against the core configuration, removing history re-anchoring or behavioral replay costs roughly 33\% of recommendation accuracy (0.049 to 0.033 and 0.034), while removing the learned routing token costs comparatively little, since routing governs when the model recommends rather than how well. That re-anchoring and replay degrade to nearly identical points despite acting through different channels, one at decoding time and one on the parameters, indicates that neither substitutes for the other. Instruction tuning data, finally, is what converts the core recipe into a usable conversational system: it leaves accuracy essentially unchanged (0.049 to 0.051) while lifting conversation quality by 0.63, and removing replay from the core recipe raises conversation quality (4.14 \(\to\) 4.37), consistent with behavioral data suppressing the language ability that rehearsal restores.

%% file: table/data_stats.tex
\begin{table}[t]
    \centering
    \caption{Dataset statistics. Stage-1 sequential recommendation data is 5-core and split by leave-one-out; \#Inter. counts every user--item event and $\overline{|H|}$ is the average length of each user history. Stage-2 conversational data is synthesized per category; a turn is one user or assistant message, $\overline{|T|}$ is the average number of turns per trajectory. Statistics are computed over all splits combined.}
    \label{tab:data}
    \setlength{\tabcolsep}{5pt}
    \resizebox{\textwidth}{!}{%
    \begin{tabular}{lrrrrcrrr}
    \toprule
    & \multicolumn{4}{c}{\textbf{Stage-1: Sequential}} & & \multicolumn{3}{c}{\textbf{Stage-2: Conversational}} \\
    \cmidrule{2-5} \cmidrule{7-9}
    \textbf{Dataset} & \textbf{\#Inter.} & \textbf{\#Users} & \textbf{\#Items} & \textbf{$\overline{|H|}$} & & \textbf{\#Turns} & \textbf{\#Traj.} & \textbf{$\overline{|T|}$} \\
    \midrule
    Beauty & 81,767 & 9,880 & 11,707 & 8.28 & & 46,594 & 6,302 & 7.39 \\
    Toys \& Games & 75,855 & 9,631 & 11,498 & 7.88 & & 39,570 & 5,647 & 7.01 \\
    \bottomrule
    \end{tabular}
    }
\end{table}

%% file: table/main.tex
\begin{table}[t]
    \centering
    \caption{Recommendation accuracy and conversation quality on Amazon Beauty and Toys. Conv.\ Q is the mean LLM-as-judge score. SASRec and TIGER are recommendation-only models, so conversation quality isn't applicable. Best results are in \textbf{bold} and second-best are \underline{underlined}.}
    \setlength{\tabcolsep}{5pt}
    \resizebox{\textwidth}{!}{%
    \begin{tabular}{lccccccc}
    \toprule
    & \multicolumn{3}{c}{\textbf{Beauty}} & & \multicolumn{3}{c}{\textbf{Toys \& Games}} \\
    \cmidrule{2-4} \cmidrule{6-8}
    \textbf{Model} & \textbf{Hits@10} & \textbf{NDCG@10} & \textbf{Conv. Q} & & \textbf{Hits@10} & \textbf{NDCG@10} & \textbf{Conv. Q} \\
    \midrule
    SASRec & 0.040 & 0.018 & -- & & 0.034 & 0.015 & -- \\
    TIGER & 0.038 & 0.020 & -- & & 0.036 & 0.020 & -- \\
    GCRS & 0.004 & 0.002 & 4.07 & & 0.002 & 0.001 & \underline{4.53} \\
    PECRS & 0.043 & 0.019 & 3.92 & & \textbf{0.057} & \textbf{0.032} & 4.05 \\
    Modular (Classic) & \underline{0.045} & 0.024 & 4.40 & & \underline{0.051} & 0.021 & 4.52 \\
    Modular (Tool-call) & 0.044 & \underline{0.025} & \underline{4.58} & & 0.049 & 0.020 & \textbf{4.55} \\
    \textbf{INTEGER (Ours)} & \textbf{0.051} & \textbf{0.027} & \textbf{4.77} & & \textbf{0.057} & \underline{0.028} & 4.25 \\
    \bottomrule
    \end{tabular}
    }
    \label{tab:main_result}
\end{table}

%% file: table/ablation.tex
\begin{table}[t]
    \centering
    \caption{Component ablation on the Amazon Beauty dataset. Removing (4) gives the core interactive recipe $(1)+(2)+(3)$; the three rows below it remove one core component from that configuration, so they are read against the $-$(4) row rather than against the full model. Each variant is evaluated in the input format it was trained on. The last row denotes trained only on the plain multi-turn conversation data. \textbf{Bold} and \underline{underline} denotes  the best and second best results, respectively.}
    \label{tab:ablation}
    \setlength{\tabcolsep}{5pt}
    \resizebox{0.8\textwidth}{!}{%
    \begin{tabular}{lccc}
    \toprule
    \textbf{Configuration} & \textbf{Hits@10} & \textbf{NDCG@10} & \textbf{Conv. Q} \\
    \midrule
    \textbf{INTEGER} & \textbf{0.051} & \textbf{0.027} & \textbf{4.77} \\
    $-$(4) instruction tuning data & \underline{0.049} & \underline{0.025} & 4.14 \\
    \midrule
    \quad $-$(1) learned routing token & 0.045 & 0.024 & 4.33 \\
    \quad $-$(2) history re-anchoring & 0.033 & 0.017 & 3.93 \\
    \quad $-$(3) behavioral replay & 0.034 & 0.018 & \underline{4.37} \\
    \midrule
    Dialogue-only finetuned & 0.029 & 0.017 & 2.39 \\
    \bottomrule
    \end{tabular}
    }
\end{table}

%% file: acl_paper/discussions.tex
\section{Discussions}
\label{sec:analysis}
\subsection{Routing Behavior}
\label{subsec:routing}
\begin{figure}[t]
    \centering
    \includegraphics[width=0.8\linewidth]{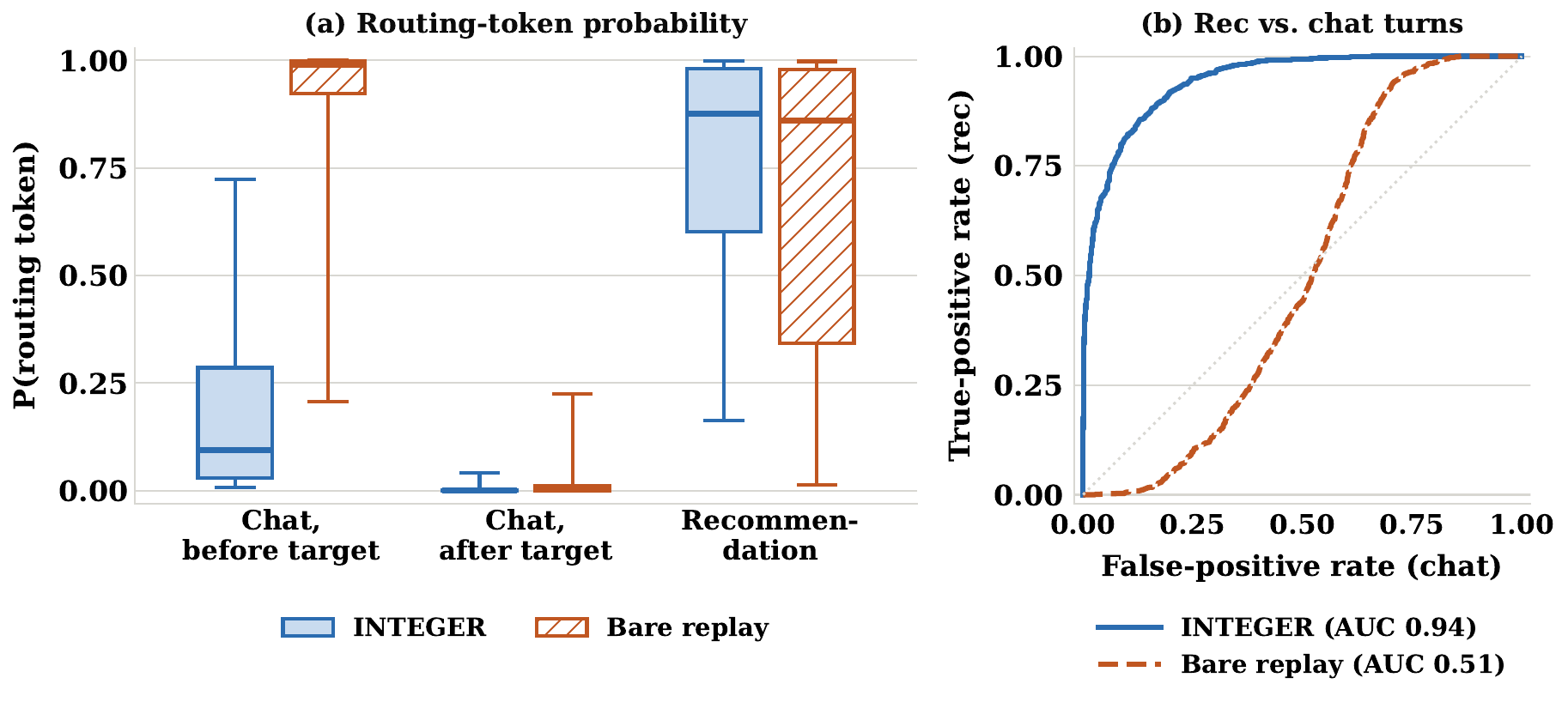}
    \caption{INTEGER knows when to recommend; bare replay does not. (a)~Probability of the routing token at each assistant turn, grouped by whether the turn should chat (before or after the target recommendation) or recommend. Well-separated boxes mean the model tells the two apart. (b)~ROC of the same probability for separating recommendation turns from chat turns.}
    \label{fig:rec_prob}
\end{figure}

Figure~\ref{fig:rec_prob} shows, at every assistant turn of the test dialogues, how strongly each model wants to recommend, i.e.\ the probability of its routing token. We compare INTEGER with \emph{bare replay}, which is trained on the same dialogues mixed with single-turn replay examples in which a recommendation is the first and only response. INTEGER separates the two kinds of turn: the median probability is $0.88$ on recommendation turns but $0.09$ on chat turns before the target recommendation and $0.00$ after it, giving an AUC of $0.94$. Bare replay's probability carries no information about the turn type (AUC $0.51$). Its median on chat turns before the target is $0.99$, higher than on recommendation turns ($0.86$).

The difference lies entirely in the turns before the target: after it, both models stay quiet on chat turns. Bare replay therefore does not over-recommend in general; it recommends before the conversation calls for it, as if a recommendation could open a dialogue regardless of what was said, which is what its single-turn replay examples teach. INTEGER wraps each replayed sequence in a dialogue, so its routing decision follows the conversational content instead: it waits while the user is still describing what they want, and recommends once they have.

\subsection{Two Directions of Forgetting}

\begin{figure}[t]
    \centering
    \includegraphics[width=0.8\linewidth]{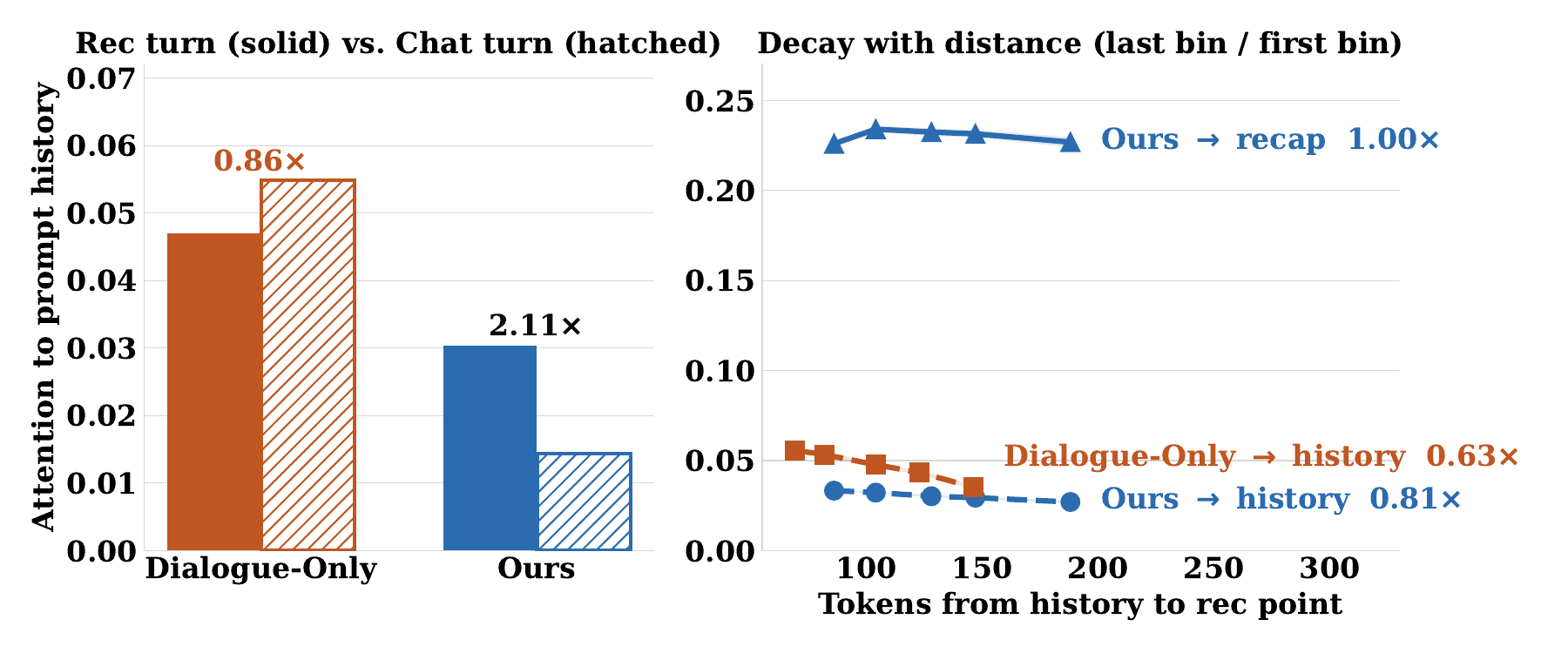}
    \caption{Left panel shows the attention score of recommendation turn against chat turn of ours and dialogue-only finetuned model. Right panel shows the attention score decay with distance.}
    \label{fig:attention}
\end{figure}

Figure~\ref{fig:attention} shows the attention to the prompted history at the moment when the model recommends an item. The model trained on dialogues alone (Table~\ref{tab:ablation}, last row) attends slightly less to the history when it recommends than when it chats (\(0.86\times\)), while ours attends \(2.11\times\) more. Focusing on prompted history at the recommendation point is a learned behavior, and dialogue-only finetuning does not acquire it. The right panel shows why an adjacent copy \(H\) helps. Attention to the far prompt history falls off as the dialogue pushes it further behind the recommendation point ($0.81\times$ for our model, $0.63\times$ for the dialogue-only model, first to last distance), while attention to the adjacent history recap \(H\) is flat ($1.00\times$). The decay is therefore not the mechanical consequence of a longer prompt, which would move all three curves together. Re-anchoring works because it puts the same tokens on a channel that dialogue length cannot dilute, and it is the copied content that matters.

\begin{figure}[t]
    \centering
    \includegraphics[width=0.8\linewidth]{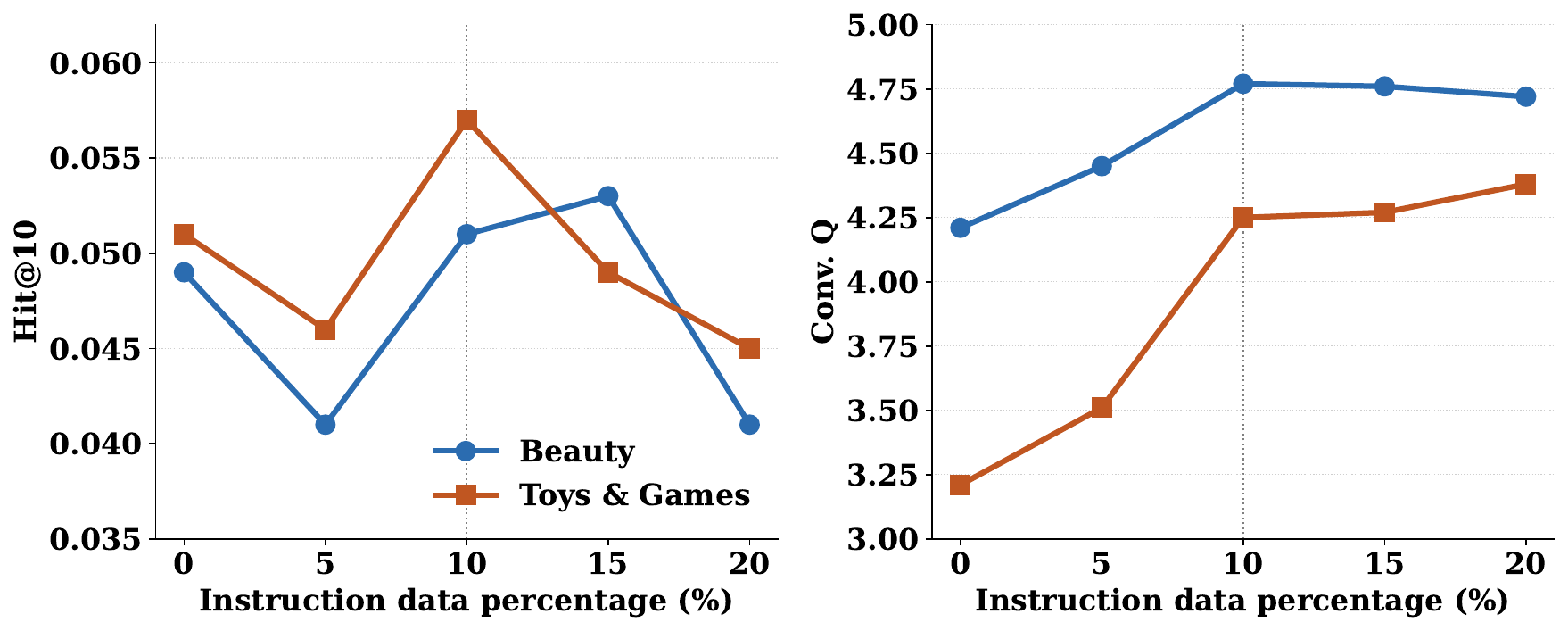}
    \caption{Recommendation performance and conversation quality over different portion of instruction tuning dataset.}
    \label{fig:dose}
\end{figure}

Figure~\ref{fig:dose} turns to the opposite direction of forgetting, reporting Hits@10 and conversation quality (Conv. Q) when adding different percentage of instruction following dataset. Conversation quality rises sharply up to 10\% and then plateaus, with an effect size of $+0.5$ to $+1.2$, while recommendation accuracy shows no systematic trend: the curve is non-monotone and its best operating point differs
by dataset. 

\subsection{Item Refinement}
\input{table/item_refine}
Table~\ref{tab:refine} reports refinement behavior under two intents, variety and rejection. The base model fails on both intents by re-offering the rejected item in every case (Repeat@10 $=1.00$, $\Delta$Rank $=0$). DPO resolves this failure, pushing the rejected item out of the top ranks (Repeat@10 $0.284$ and $0.269$; $\Delta$Rank $22$ and $26$) and nearly doubling recovery of the gold item ($0.228 \rightarrow 0.367$ and $0.182 \rightarrow 0.364$).
Sim.\ measures how far the new recommendation moves from the rejected one, and a lower value is not inherently better. A variety request calls for an item that remains close to the current one, whereas a rejection may call for moving further away. Yet DPO lands at the same distance under both intents ($0.885$ and $0.886$), and the two intents also match closely on every other measure. This pattern follows from how the preference pairs are constructed. Both pair types reward replacing the current item and neither carries an intent label, so the model learns a single replacement operation over the item space rather than intent-specific behaviors, even though the two intents are expressed by distinct surface phrasings.

\subsection{Case Study}
\begin{figure}[t]
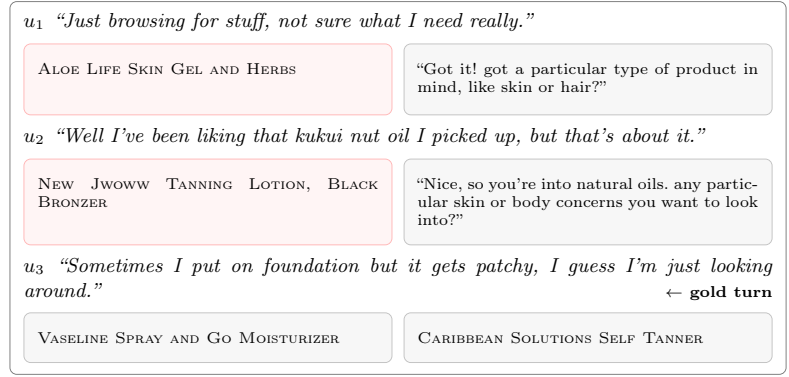

\centering
\resizebox{0.85\textwidth}{!}{%
\begin{tcolorbox}[
  colback=white, colframe=black!55, boxrule=0.4pt, left=3pt, right=3pt, top=2pt, bottom=2pt]
\footnotesize

$u_1$\, \emph{``Just browsing for stuff, not sure what I need really.''}
\begin{tcbraster}[raster columns=2, raster column skip=5pt, raster equal height=rows]
  \begin{tcolorbox}[colback=red!4, colframe=red!30, boxrule=0.4pt, left=3pt, right=3pt]
  \scriptsize \textsc{Aloe Life Skin Gel and Herbs}
  \end{tcolorbox}
  \begin{tcolorbox}[colback=black!3, colframe=black!25, boxrule=0.4pt, left=3pt, right=3pt]
  \scriptsize ``Got it! got a particular type of product in mind, like skin or hair?''
  \end{tcolorbox}
\end{tcbraster}

$u_2$\, \emph{``Well I've been liking that kukui nut oil I picked up, but that's about it.''}
\begin{tcbraster}[raster columns=2, raster column skip=5pt, raster equal height=rows]
  \begin{tcolorbox}[colback=red!4, colframe=red!30, boxrule=0.4pt, left=3pt, right=3pt]
  \scriptsize \textsc{New Jwoww Tanning Lotion, Black Bronzer}
  \end{tcolorbox}
  \begin{tcolorbox}[colback=black!3, colframe=black!25, boxrule=0.4pt, left=3pt, right=3pt]
  \scriptsize ``Nice, so you're into natural oils. any particular skin
  or body concerns you want to look into?''
  \end{tcolorbox}
\end{tcbraster}

$u_3$\, \emph{``Sometimes I put on foundation but it gets patchy, I guess I'm
just looking around.''}\hfill\textbf{\scriptsize$\leftarrow$ gold turn}
\begin{tcbraster}[raster columns=2, raster column skip=5pt, raster equal height=rows]
  \begin{tcolorbox}[colback=black!3, colframe=black!25, boxrule=0.4pt, left=3pt, right=3pt]
  \scriptsize \textsc{Vaseline Spray and Go Moisturizer}
  \end{tcolorbox}
  \begin{tcolorbox}[colback=black!3, colframe=black!25, boxrule=0.4pt, left=3pt, right=3pt]
  \scriptsize \textsc{Caribbean Solutions Self Tanner}
  \end{tcolorbox}
\end{tcbraster}
\end{tcolorbox}
}
\caption{Case study on Amazon Beauty dialogue: bare-replay (left) recommends prematurely at every turn (\textcolor{red!30}{red}), while INTEGER (right) asks clarifying questions and recommends only at the gold turn.}
\label{fig:case}
\end{figure}

Figure~\ref{fig:case} illustrates the routing difference from Section~\ref{subsec:routing} on a single dialogue. The user opens without a clear goal \emph{"just browsing"} and only gradually reveals what they want. Bare replay recommends an item at every assistant turn, even before the user has expressed any preference. INTEGER instead responds conversationally, asking about product type and skin concerns, and withholds a recommendation until the third turn, when the user mentions patchy foundation.

%% file: table/item_refine.tex
\begin{table}[t]
    \centering
    \caption{Item refinement on recoverable reject scenarios on Amazon Beauty, where the gold item was originally ranked $2$--$50$ (each value averages three surface phrasings of the intent). Suppression measured with Repeat@10 and $\Delta$Rank of the rejected item applies to both intents. Alternative reports the embedding similarity (Sim.) between the re-offered item and the rejected one (lower is more distinct), and Recovery reports Golden Hit@10, whether the originally ranked gold item is recovered. The base model re-offers the rejected item under both intents, so its alternative is near-identical and $\Delta$Rank is zero.}
    \label{tab:refine}
    \setlength{\tabcolsep}{5pt}
    \resizebox{0.8\textwidth}{!}{%
    \begin{tabular}{llcccc}
    \toprule
    & & \multicolumn{2}{c}{\textbf{Suppression}} & \textbf{Alternative} & \textbf{Recovery} \\
    \cmidrule(lr){3-4}
    \textbf{Intent} & & Repeat@10 & $\Delta$Rank & Sim. & Golden Hit@10 \\
    \midrule
    \multirow{2}{*}{Variety}
     & Base & 1.000 & 0 & 0.997 & 0.228 \\
     & \ \emph{\textbf{+DPO}} & 0.284 & 22 & 0.885 & 0.367 \\
    \midrule
    \multirow{2}{*}{Rejection}
     & Base & 1.000 & 0 & 0.995 & 0.182 \\
     & \ \emph{\textbf{+DPO}} & 0.269 & 26 & 0.886 & 0.364 \\
    \bottomrule
    \end{tabular}
    }
\end{table}

%% file: acl_paper/conclusions_and_future_works.tex
\section{Conclusion \& Future Work}
\label{sec:conclusion}
In this work, we presented INTEGER, a model that extends a generative recommender to multi-turn interaction by exposing recommendation as a decoding action rather than learning it from dialogue. On Amazon Beauty and Toys, INTEGER is competitive with both end-to-end conversational recommenders and modular pipelines in accuracy and conversation quality, and significantly outperforms the generative recommender it starts from. Our analyses show that replay, re-anchoring, and rehearsal together prevent conversational and behavioral training from eroding each other. Our findings are limited by the use of synthesized dialogues and by refinement that recovers near-misses rather than satisfying stated preferences. Handling attribute-level feedback (e.g., \emph{too expensive}, \emph{I want a black one}) would require conditioning refinement on item attributes, whereas the operation we learn is intent-agnostic and moves only within the embedding neighborhood.

%% file: ecir_preprint.bib
@article{yang2025qwen3,
  title={Qwen3 technical report},
  author={Yang, An and Li, Anfeng and Yang, Baosong and Zhang, Beichen and Hui, Binyuan and Zheng, Bo and Yu, Bowen and Gao, Chang and Huang, Chengen and Lv, Chenxu and others},
  journal={arXiv preprint arXiv:2505.09388},
  year={2025}
}

@article{rajput2023recommender,
  title={Recommender systems with generative retrieval},
  author={Rajput, Shashank and Mehta, Nikhil and Singh, Anima and Hulikal Keshavan, Raghunandan and Vu, Trung and Heldt, Lukasz and Hong, Lichan and Tay, Yi and Tran, Vinh and Samost, Jonah and others},
  journal={Advances in Neural Information Processing Systems},
  volume={36},
  pages={10299--10315},
  year={2023}
}

@inproceedings{he2026plum,
  title={Plum: Adapting pre-trained language models for industrial-scale generative recommendations},
  author={He, Ruining and Heldt, Lukasz and Hong, Lichan and Keshavan, Raghunandan and Mao, Shifan and Mehta, Nikhil and Su, Zhengyang and Tsai, Alicia and Wang, Yueqi and Wang, Shao-Chuan and others},
  booktitle={Proceedings of the ACM Web Conference 2026},
  pages={8093--8104},
  year={2026}
}

@inproceedings{zhai2024actions,
  title={Actions Speak Louder than Words: Trillion-Parameter Sequential Transducers for Generative Recommendations},
  author={Zhai, Jiaqi and Liao, Lucy and Liu, Xing and Wang, Yueming and Li, Rui and Cao, Xuan and Gao, Leon and Gong, Zhaojie and Gu, Fangda and He, Jiayuan and others},
  booktitle={International Conference on Machine Learning},
  pages={58484--58509},
  year={2024},
  organization={PMLR}
}

@inproceedings{han2025mtgr,
  title={Mtgr: Industrial-scale generative recommendation framework in meituan},
  author={Han, Ruidong and Yin, Bin and Chen, Shangyu and Jiang, He and Jiang, Fei and Li, Xiang and Ma, Chi and Huang, Mincong and Li, Xiaoguang and Jing, Chunzhen and others},
  booktitle={Proceedings of the 34th ACM International Conference on Information and Knowledge Management},
  pages={5731--5738},
  year={2025}
}

@inproceedings{ju2025generative,
  title={Generative Recommendation with Semantic IDs: A Practitioner's Handbook},
  author={Ju, Clark Mingxuan and Collins, Liam and Neves, Leonardo and Kumar, Bhuvesh and Wang, Louis Yufeng and Zhao, Tong and Shah, Neil},
  booktitle={Proceedings of the 34th ACM International Conference on Information and Knowledge Management},
  pages={6420--6425},
  year={2025}
}

@inproceedings{zhu2024cost,
  title={Cost: Contrastive quantization based semantic tokenization for generative recommendation},
  author={Zhu, Jieming and Jin, Mengqun and Liu, Qijiong and Qiu, Zexuan and Dong, Zhenhua and Li, Xiu},
  booktitle={Proceedings of the 18th ACM Conference on Recommender Systems},
  pages={969--974},
  year={2024}
}

@article{zhang2026generative,
  title={Generative Conversational Recommender System},
  author={Zhang, Sixiao and Liu, Mingrui and Long, Cheng},
  journal={arXiv preprint arXiv:2605.21987},
  year={2026}
}

@inproceedings{chen2025llm,
  title={LLM-Based Dialogue Labeling for Multiturn Adaptive RAG},
  author={Chen, Zhiyu and Xie, Biancen and Srinivasan, Sidarth and Ramanathan, Manikandarajan and Maragoud, Rajashekar and Liu, Qun},
  booktitle={Proceedings of the 2025 Conference on Empirical Methods in Natural Language Processing: Industry Track},
  pages={1044--1056},
  year={2025}
}

@article{pattnayak2025hybrid,
  title={Hybrid AI for Responsive Multi-Turn Online Conversations with Novel Dynamic Routing and Feedback Adaptation},
  author={Pattnayak, P and others},
  journal={arXiv preprint arXiv:2506.02097},
  year={2025}
}

@inproceedings{papineni2002bleu,
  title={Bleu: a method for automatic evaluation of machine translation},
  author={Papineni, Kishore and Roukos, Salim and Ward, Todd and Zhu, Wei-Jing},
  booktitle={Proceedings of the 40th annual meeting of the Association for Computational Linguistics},
  pages={311--318},
  year={2002}
}

@inproceedings{lin2004rouge,
  title={Rouge: A package for automatic evaluation of summaries},
  author={LIN, CY},
  booktitle={Text Summarization Branches Out: Proceedings of the ACL-04 Workshop, Barcelona, Spain},
  pages={74--81},
  year={2004}
}

@article{deng2025onerec,
  title={Onerec: Unifying retrieve and rank with generative recommender and iterative preference alignment},
  author={Deng, Jiaxin and Wang, Shiyao and Cai, Kuo and Ren, Lejian and Hu, Qigen and Ding, Weifeng and Luo, Qiang and Zhou, Guorui},
  journal={arXiv preprint arXiv:2502.18965},
  year={2025}
}

@inproceedings{hou2026bridging,
  title={Bridging language and items for retrieval and recommendation: Benchmarking LLMs as semantic encoders},
  author={Hou, Yupeng and Li, Jiacheng and Fu, Xiangjun and He, Zhankui and Yan, An and Chen, Xiusi and McAuley, Julian},
  booktitle={Proceedings of the 64th Annual Meeting of the Association for Computational Linguistics (Volume 1: Long Papers)},
  pages={3251--3265},
  year={2026}
}

@inproceedings{kang2018self,
  title={Self-attentive sequential recommendation},
  author={Kang, Wang-Cheng and McAuley, Julian},
  booktitle={2018 IEEE international conference on data mining (ICDM)},
  pages={197--206},
  year={2018},
  organization={IEEE}
}

@inproceedings{sun2019bert4rec,
  title={BERT4Rec: Sequential recommendation with bidirectional encoder representations from transformer},
  author={Sun, Fei and Liu, Jun and Wu, Jian and Pei, Changhua and Lin, Xiao and Ou, Wenwu and Jiang, Peng},
  booktitle={Proceedings of the 28th ACM international conference on information and knowledge management},
  pages={1441--1450},
  year={2019}
}

@article{hidasi2015session,
  title={Session-based recommendations with recurrent neural networks},
  author={Hidasi, Bal{\'a}zs and Karatzoglou, Alexandros and Baltrunas, Linas and Tikk, Domonkos},
  journal={arXiv preprint arXiv:1511.06939},
  year={2015}
}

@inproceedings{ma2019hierarchical,
  title={Hierarchical gating networks for sequential recommendation},
  author={Ma, Chen and Kang, Peng and Liu, Xue},
  booktitle={Proceedings of the 25th ACM SIGKDD international conference on knowledge discovery \& data mining},
  pages={825--833},
  year={2019}
}

@inproceedings{liu2025generative,
  title={Generative recommender with end-to-end learnable item tokenization},
  author={Liu, Enze and Zheng, Bowen and Ling, Cheng and Hu, Lantao and Li, Han and Zhao, Wayne Xin},
  booktitle={Proceedings of the 48th International ACM SIGIR Conference on Research and Development in Information Retrieval},
  pages={729--739},
  year={2025}
}

@article{vandenhirtz2026multimodal,
  title={Multimodal Generative Recommendation for Fusing Semantic and Collaborative Signals},
  author={Vandenhirtz, Moritz and Hassani, Kaveh and Ghasemlou, Shervin and Shao, Shuai and Eghbalzadeh, Hamid and Peng, Fuchun and Liu, Jun and Iuzzolino, Michael Louis},
  journal={arXiv preprint arXiv:2602.03713},
  year={2026}
}

@article{liu2024mmgrec,
  title={Mmgrec: Multimodal generative recommendation with transformer model},
  author={Liu, Han and Wei, Yinwei and Song, Xuemeng and Guan, Weili and Li, Yuan-Fang and Nie, Liqiang},
  journal={arXiv preprint arXiv:2404.16555},
  year={2024}
}

@inproceedings{xu2026mmq,
  title={Mmq: Multimodal mixture-of-quantization tokenization for semantic id generation and user behavioral adaptation},
  author={Xu, Yi and Zhang, Moyu and Li, Chenxuan and Liao, Zhihao and Xing, Haibo and Deng, Hao and Hu, Jinxin and Zhang, Yu and Zeng, Xiaoyi and Zhang, Jing},
  booktitle={Proceedings of the Nineteenth ACM International Conference on Web Search and Data Mining},
  pages={788--797},
  year={2026}
}

@article{zhu2025beyond,
  title={Beyond unimodal boundaries: Generative recommendation with multimodal semantics},
  author={Zhu, Jing and Ju, Mingxuan and Liu, Yozen and Koutra, Danai and Shah, Neil and Zhao, Tong},
  journal={arXiv preprint arXiv:2503.23333},
  year={2025}
}

@inproceedings{hou2025generating,
  title={Generating long semantic ids in parallel for recommendation},
  author={Hou, Yupeng and Li, Jiacheng and Shin, Ashley and Jeon, Jinsung and Santhanam, Abhishek and Shao, Wei and Hassani, Kaveh and Yao, Ning and McAuley, Julian},
  booktitle={Proceedings of the 31st ACM SIGKDD Conference on Knowledge Discovery and Data Mining V. 2},
  pages={956--966},
  year={2025}
}

@article{zheng2025pre,
  title={Pre-training generative recommender with multi-identifier item tokenization},
  author={Zheng, Bowen and Liu, Enze and Chen, Zhongfu and Ma, Zhongrui and Wang, Yue and Zhao, Wayne Xin and Wen, Ji-Rong},
  journal={arXiv preprint arXiv:2504.04400},
  year={2025}
}

@inproceedings{wang2026nezha,
  title={Nezha: A zero-sacrifice and hyperspeed decoding architecture for generative recommendations},
  author={Wang, Yejing and Zhou, Shengyu and Lu, Jinyu and Liu, Ziwei and Liu, Langming and Wang, Maolin and Zhang, Wenlin and Li, Feng and Su, Wenbo and Wang, Pengjie and others},
  booktitle={Proceedings of the ACM Web Conference 2026},
  pages={8073--8082},
  year={2026}
}

@inproceedings{lin2026bringing,
  title={Bringing reasoning to generative recommendation through the lens of cascaded ranking},
  author={Lin, Xinyu and Liu, Pengyuan and Wang, Wenjie and Hu, Yicheng and Xu, Chen and Feng, Fuli and Wang, Qifan and Chua, Tat-Seng},
  booktitle={Proceedings of the ACM Web Conference 2026},
  pages={8939--8949},
  year={2026}
}

@inproceedings{ravaut2024parameter,
  title={Parameter-efficient conversational recommender system as a language processing task},
  author={Ravaut, Mathieu and Zhang, Hao and Xu, Lu and Sun, Aixin and Liu, Yong},
  booktitle={Proceedings of the 18th Conference of the European Chapter of the Association for Computational Linguistics (Volume 1: Long Papers)},
  pages={152--165},
  year={2024}
}

@inproceedings{tajiri2025refining,
  title={Refining text generation for realistic conversational recommendation via direct preference optimization},
  author={Tajiri, Manato and Inaba, Michimasa},
  booktitle={Proceedings of the 2025 Conference on Empirical Methods in Natural Language Processing},
  pages={28628--28649},
  year={2025}
}

@inproceedings{wang2022towards,
  title={Towards unified conversational recommender systems via knowledge-enhanced prompt learning},
  author={Wang, Xiaolei and Zhou, Kun and Wen, Ji-Rong and Zhao, Wayne Xin},
  booktitle={Proceedings of the 28th ACM SIGKDD conference on knowledge discovery and data mining},
  pages={1929--1937},
  year={2022}
}

@article{verma2026orbit,
  title={ORBIT: Preserving Foundational Language Capabilities in GenRetrieval via Origin-Regulated Merging},
  author={Verma, Neha and Mehta, Nikhil and Wang, Shao-Chuan and Zhang, Naijing and Tsai, Alicia and Wei, Li and Heldt, Lukasz and Hong, Lichan and Chi, Ed and Yi, Xinyang},
  journal={arXiv preprint arXiv:2605.12419},
  year={2026}
}

@article{rafailov2023direct,
  title={Direct preference optimization: Your language model is secretly a reward model},
  author={Rafailov, Rafael and Sharma, Archit and Mitchell, Eric and Manning, Christopher D and Ermon, Stefano and Finn, Chelsea},
  journal={Advances in neural information processing systems},
  volume={36},
  pages={53728--53741},
  year={2023}
}

@inproceedings{lee2022autoregressive,
  title={Autoregressive image generation using residual quantization},
  author={Lee, Doyup and Kim, Chiheon and Kim, Saehoon and Cho, Minsu and Han, Wook-Shin},
  booktitle={2022 IEEE/CVF conference on computer vision and pattern recognition (CVPR)},
  pages={11513--11522},
  year={2022},
  organization={IEEE}
}

@article{jegou2010product,
  title={Product quantization for nearest neighbor search},
  author={Jegou, Herve and Douze, Matthijs and Schmid, Cordelia},
  journal={IEEE transactions on pattern analysis and machine intelligence},
  volume={33},
  number={1},
  pages={117--128},
  year={2010},
  publisher={IEEE}
}

@inproceedings{kim2025lost,
  title={Lost in sequence: Do large language models understand sequential recommendation?},
  author={Kim, Sein and Kang, Hongseok and Kim, Kibum and Kim, Jiwan and Kim, Donghyun and Yang, Minchul and Oh, Kwangjin and McAuley, Julian and Park, Chanyoung},
  booktitle={Proceedings of the 31st ACM SIGKDD Conference on Knowledge Discovery and Data Mining V. 2},
  pages={1160--1171},
  year={2025}
}

@inproceedings{wang2025cesrec,
  title={CESRec: Constructing Pseudo Interactions for Sequential Recommendation via Conversational Feedback.},
  author={Wang, Yifan and Gao, Shen and Fang, Jiabao and Yan, Rui and Chiu, Billy and Shang, Shuo},
  booktitle={EMNLP (Findings)},
  pages={16227--16239},
  year={2025}
}

@article{xu2026deepseek,
  title={Deepseek-v4: Towards highly efficient million-token context intelligence},
  author={Xu, Anyi and Lin, Bangcai and Xue, Bing and Wang, Bingxuan and Xu, Bingzheng and Wu, Bochao and Zhang, Bowei and Lin, Chaofan and Dong, Chen and Ling, Chenchen and others},
  journal={arXiv preprint arXiv:2606.19348},
  year={2026}
}

@inproceedings{ding2023enhancing,
  title={Enhancing chat language models by scaling high-quality instructional conversations},
  author={Ding, Ning and Chen, Yulin and Xu, Bokai and Qin, Yujia and Hu, Shengding and Liu, Zhiyuan and Sun, Maosong and Zhou, Bowen},
  booktitle={Proceedings of the 2023 Conference on Empirical Methods in Natural Language Processing},
  pages={3029--3051},
  year={2023}
}

@misc{tencent2025hy3preview,
  author       = {Tencent Hunyuan Team},
  title        = {Hy3-preview},
  year         = {2025},
  publisher    = {Hugging Face},
  journal      = {Hugging Face Repository},
  howpublished = {\url{https://huggingface.co/tencent/Hy3-preview}}
}

@article{gao2023chatrec,
  title={Chat-REC: Towards Interactive and Explainable LLMs-Augmented Recommender System},
  author={Gao, Yunfan and Sheng, Tao and Xiang, Youlin and Xiong, Yun and Wang, Haofen and Zhang, Jiawei},
  journal={arXiv preprint arXiv:2303.14524},
  year={2023},
  doi={10.48550/arXiv.2303.14524}
}

@inproceedings{zhao2024toolrec,
  title={Let Me Do It For You: Towards LLM Empowered Recommendation via Tool Learning},
  author={Zhao, Yuyue and Wu, Jiancan and Wang, Xiang and Tang, Wei and Wang, Dingxian and de Rijke, Maarten},
  booktitle={Proceedings of the 47th International ACM SIGIR Conference on Research and Development in Information Retrieval},
  year={2024},
  doi={10.1145/3626772.3657828}
}

@article{huang2025interecagent,
  title={Recommender AI Agent: Integrating Large Language Models for Interactive Recommendations},
  author={Huang, Xu and Lian, Jianxun and Lei, Yuxuan and Yao, Jing and Lian, Defu and Xie, Xing},
  journal={ACM Transactions on Information Systems},
  volume={43},
  number={4},
  pages={1--33},
  year={2025},
  doi={10.1145/3731446}
}

@inproceedings{li2025chatcrs,
  title={ChatCRS: Incorporating External Knowledge and Goal Guidance for LLM-based Conversational Recommender Systems},
  author={Li, Chuang and Deng, Yang and Hu, Hengchang and Kan, Min-Yen and Li, Haizhou},
  booktitle={Findings of the Association for Computational Linguistics: NAACL 2025},
  pages={295--312},
  year={2025},
  publisher={Association for Computational Linguistics},
  doi={10.18653/v1/2025.findings-naacl.17}
}

@inproceedings{ni2022sentence,
  title={Sentence-t5: Scalable sentence encoders from pre-trained text-to-text models},
  author={Ni, Jianmo and Abrego, Gustavo Hernandez and Constant, Noah and Ma, Ji and Hall, Keith and Cer, Daniel and Yang, Yinfei},
  booktitle={Findings of the association for computational linguistics: ACL 2022},
  pages={1864--1874},
  year={2022}
}

@article{yang2026sparse,
  title={Sparse meets dense: Unified generative recommendations with cascaded sparse-dense representations},
  author={Yang, Yuhao and Ji, Zhi and Li, Zhaopeng and Li, Yi and Mo, Zhonglin and Ding, Yue and Chen, Kai and Zhang, Zijian and Li, Jie and LIN, LIU and others},
  journal={Advances in Neural Information Processing Systems},
  volume={38},
  pages={93746--93770},
  year={2026}
}

@inproceedings{doh2026llm,
author = {Doh, Seungheon and Sguerra, Bruno and Oramas, Sergio and Epure, Elena V. and Nam, Juhan},
title = {LLM-as-a-Judge for Evaluating System Responses in Conversational Music Recommendation},
year = {2026},
isbn = {9798400722844},
publisher = {Association for Computing Machinery},
address = {New York, NY, USA},
booktitle = {Proceedings of the 20th ACM Conference on Recommender Systems},
pages = {708–714},
numpages = {7},
location = {Minneapolis, USA},
series = {RecSys '26}
}

@article{zheng2023judging,
  title={Judging llm-as-a-judge with mt-bench and chatbot arena},
  author={Zheng, Lianmin and Chiang, Wei-Lin and Sheng, Ying and Zhuang, Siyuan and Wu, Zhanghao and Zhuang, Yonghao and Lin, Zi and Li, Zhuohan and Li, Dacheng and Xing, Eric and others},
  journal={Advances in neural information processing systems},
  volume={36},
  pages={46595--46623},
  year={2023}
}
